\documentclass[a4paper]{amsart}

\date{March 19, 2013}

\newtheorem{theorem}{Theorem}

\def\rz{\mathbb{R}} % reelle Zahlen
\def\cF{\mathcal{F}}
\def\cH{\mathcal{H}}
\def\cJ{\mathcal{J}}

\def\gp{\mathfrak{p}}
\def\gq{\mathfrak{q}}

\def\rd{\mathrm{d}}
\def\ri{\mathrm{i}}

\begin{document}
\title[$0\leq|\gp|^a|\gq|^b+|\gq|^b|\gp|^a$]{Positivity of $|\gp|^a|\gq|^b+|\gq|^b|\gp|^a$}

\author[L. Chen]{Li Chen}
\address{Department of Mathematical Sciences\\ Tsinghua University\\ Beijing 100084\\ China}
\author[H. Siedentop]{Heinz Siedentop}
\address{Mathematisches Institut\\
 Ludwig-Maximilians-Universit\"at M\"unchen\\
 Theresienstra\ss e 39\\ 80333 M\"unchen\\ Germany}
 \email{lchen@math.tsinghua.edu.cn \textrm{and} h.s@lmu.de}

\maketitle

\begin{abstract}
  We show that
  $$\cJ_{a,b,n}:=\frac12(|\gp|^a|\gq|^b+|\gq|^b|\gp|^a)$$
  is positive under suitable conditions on the exponents $a$ and $b$
  and the underlying dimension $n$. (Here $\gq$ is the multiplication
  by $x$ and $\gp:= \mathrm{i}^{-1}\nabla$.) Furthermore we show a
  generalization of the generalized Hardy inequalities for the
  fractional Laplacians.
 \end{abstract}

\section{Introduction\label{ein}}

The classical Hardy inequality (\cite[Formula (4)]{Hardy1920}))
$$\int_a^\infty\left(F\over x\right)^\kappa \rd x \leq \left(\kappa\over\kappa-1\right)^\kappa\int_a^\infty f^\kappa \rd x$$ 
with $F(x)=\int_a^xf(t)\rd t$ and $\kappa>1$ is one of the longest
known inequalities allowing to bound the weighted $L^\kappa$-norm of a
decaying function by the $L^\kappa$-norm of its gradient (Hardy
\cite{Hardy1920}). In modern textbooks, see, e.g., Reed and Simon
\cite[p. 169]{ReedSimon1975}, this occurs ($\kappa=2$) as the quantum
mechanical uncertainty principle lemma and is written in three
dimensions as
\begin{equation}
  \label{eq:uncertain}
  \int_{\rz^3} |\nabla\psi|^2 \geq \frac14 \int_{\rz^3} {|\psi(x)|^2\over |x|^2}\rd x.
  \end{equation}
  This, in turn was generalized by Herbst \cite{Herbst1977} (see also
  Yafaev \cite{Yafaev1999} and Frank et al \cite{Franketal2008H}) to
  fractional Laplacians (see \eqref{eq:hardy}).

  In a seemingly different context, the excess charge problem of
  atoms, Lieb\cite{Lieb1984} needed
\begin{eqnarray}\label{Lieb}
|\gq||\gp|^2+|\gp|^2|\gq|> 0
\end{eqnarray}
which, however, turned out to be equivalent to the quantum mechanical
uncertainty principle. Here $\gp = -\ri \nabla$ is the momentum
operator and $\gq$ (multiplication by $x$) is the position
operator. Lieb \cite{Lieb1984} showed in fact, that also
\begin{equation}
  \label{liebrel}
  |\gq||\gp|+|\gp||\gq|> 0
\end{equation}
in three dimensions by reducing it to \eqref{Lieb}.

With the advent of graphene physics, two-dimensional versions of
Lieb's inequality became of physical interest which, however, could
not simply be reduced to \eqref{Lieb}. Instead,
\eqref{liebrel} was directly proven \cite{HandrekSiedentop2013}.

The purpose of this article is to show, that the positivity of the
Jordan product $\cJ_{a,b,n}:=\frac12(|\gp|^a|\gq|^b+|\gq|^b|\gp|^a)$
is in fact a generalization which reduces, for $b=n-a$ to Hardy
inequalities for fractional Laplacians. Here $a$ and $b$ are positive
constants and $n$ is the underlying dimension of the appropriate
function space.

\section{Positivity and Relation to Generalized Hardy Inequalities\label{s:1}}

Our basic result is the following operator inequality on $L^2(\rz^n)$
for the momentum operator $\gp = -\ri \nabla$ and the position
operator $\gq$ (multiplication by $x$).
\begin{theorem}
  Assume $n\geq a+b$ and $\min\{a,b\}\in[0,2]$. Then on $C_0^\infty(\rz^n)$
  \begin{equation}
    \label{grund}
    0<\cJ_{a,b,n}:=\frac12(|\gp|^a|\gq|^b+|\gq|^b|\gp|^a).
  \end{equation}
 \end{theorem}
In fact, our proof shows more, namely
\begin{equation}
  \label{eq:mehr}
  |\gq|^{b/2}\cH_{a,n}|\gq|^{b/2}\leq \cJ_{a,b,n}
\end{equation}
where $\cH_{a,n}$ is the Hardy operator of \eqref{eq:hardy}.

As indicated in the introduction, the case $n=3$, $a=2$, and $b=1$ has
an important consequence in atomic physics: it is an essential
ingredient in bounding the total number of electrons that atoms can
bind: the number of electrons that an atom can bind can never exceed
twice its nuclear charge. This special case was proven and applied in
this context by Lieb \cite{Lieb1984}. The case $n=2$ and $a=b=1$ plays
a similar role in investigating how many electrons a magnetic quantum
dot in a graphene layer can bound and was proved and applied in that
context (Handrek and Siedentop \cite{HandrekSiedentop2013}).
\begin{proof}
  For the proof we can assume that $a\leq b$, since, if not, we use the
  Fourier transform to exchange the role of $\gp$ and $\gq$.

  We first treat the case, that $a<2$. In this case we follow the
  strategy of \cite{HandrekSiedentop2013} and use the identity
  \eqref{eq:betrag}. Thus, by polarization
  \begin{equation}
    \label{eq:4}
    \begin{split}
    t:=  &\frac12(\psi,(|\gp|^a|\gq|^b+|\gq|^b|\gp|^a)\psi) \\
    = &\alpha_{a,n}\Re\int_{\rz^n}\rd x \int_{\rz^n}\rd y
    {(\overline{\psi(x)}-\overline{\psi(y)})(|x|^b\psi(x)-|y|^b\psi(y))\over|x-y|^{n+a}}\\
    =& \alpha_{a,n}\Re \int_{\rz^n} \rd x\int_{\rz^n}\rd y
    {2|x|^b |\psi(x)|^2 - (|x|^b +|y|^b){\overline{\psi(x)}}\psi(y)\over
        |x-y|^{n+a}}.
  \end{split}
  \end{equation}

  Now, setting $\psi= g / |.|^{(n+b-a)/2}$ we get
\begin{equation}
  \label{bound}
  \begin{split}
    {t\over\alpha_{a,n}} =&\int_{\rz^n} \rd x \int_{\rz^n} \rd y  {{|x|^b|g(x)|^2\over|x|^{n+b-a}}+{|y|^b|g(y)|^2\over|y|^{n+b-a}}-{2\Re \overline{g(x)}g(y)|y|^b\over|x|^{(n+b-a)/2} |y|^{(n+b-a)/2}} \over |x-y|^{n+a}}\\
    =&\int_{\rz^n} {\rd x\over|x|^n} |g(x)|^2 \int_{\rz^n} \rd y
    {2|x|^a- |x|^{n+a-b\over2} |y|^{-n+a+b\over2} -|x|^{n+a+b\over2}|y|^{-n+a-b\over2}\over|x-y|^{n+a}}\\
    &+ \int_{\rz^n} \rd x \int_{\rz^n} \rd y
    {(|x|^b+|y|^b)|g(x)-g(y)|^2\over
      2|x|^{(n+b-a)/2}|x-y|^{n+a}|y|^{(n+b-a)/2}}.
  \end{split}
  \end{equation}
  At this point we could simply drop the last term, since it is
  positive. However, with minimal extra effort we estimate the last
  term using $|x|^b+|y|^b \geq 2 |x|^{b/2}|y|^{b/2}$ and obtain using \eqref{eq:groundstatetransformedhardy}
\begin{equation}
  \label{boundweiter}
  \begin{split}
   t 
\geq & \alpha_{a,n}\int_{\rz^n} \rd x {|g(x)|^2\over|x|^n} \int_{\rz^n}\rd y
    {2 -|y|^{a+b-n\over2} -|y|^{-n+a-b\over2}\over (2|y|)^{n+a\over2}
      \left({|y|+|y|^{-1}\over2} - \omega\cdot \mathfrak{e}\right)^{{n+a\over2}}}+(\psi,|\gq|^{b/2}\cH_{a,n}|\gq|^{b/2}\psi)\\
    =& { \alpha_{a,n}\over2^{n+a\over2}}\int_{\rz^n} \rd x {|g(x)|^2\over|x|^n}
    \int_{\rz^n} {\rd y \over |y|^n}{
    2|y|^{n-a\over2} -|y|^{b/2}-|y|^{-b/2} \over
      ({|y|+|y|^{-1}\over2} - \omega\cdot \mathfrak{e})^{{n+a\over2}}}+(\psi,|\gq|^{b/2}\cH_{a,n}|\gq|^{b/2}\psi)\\
    = &{\alpha_{a,n}\over2^{n+a\over2}}\int_{\rz^n} \rd x {|g(x)|^2\over|x|^n}
    \int_{\rz^n} {\rd y \over |y|^n}{
    |y|^{n-a\over2} +|y|^{-{n-a\over2}} -|y|^{b\over2}-|y|^{-{b\over2}} \over
      ({|y|+|y|^{-1}\over2} - \omega\cdot \mathfrak{e})^{{n+a\over2}}}+(\psi,|\gq|^{b/2}\cH_{a,n}|\gq|^{b/2}\psi)\\
>&0
    \end{split}
\end{equation}
assuming -- in the last line -- that $\psi$ is not identical zero. The
positivity, i.e., the last inequality, follows from the positivity of
the numerator of the last integral which is a consequence of the fact
that the function $f(\alpha) := r^\alpha +r^{-\alpha}$ is strictly
monotone increasing for positive $r$ and $n-a\geq b$.

We now supply the missing case that $\min\{a,b\}=2$. Again we may
assume that $a\leq b$ without loss of generality. An easy calculation
shows
\begin{equation}
  \label{giveanumber}
  \begin{split}
  &\frac12(|\gp|^2|\gq|^b+|\gq|^b|\gp|^2)\\
  =& |\gq|^{\frac{b}{2}}\Big( |\gp|^2 + \dfrac{1}{2}|\gq|^{-\frac{b}{2}}[[|\gp|^2,|\gq|^{\frac{b}{2}}],|\gq|^{\frac{b}{2}}]|\gq|^{-\frac{b}{2}}\Big)   |\gq|^{\frac{b}{2}}\\
=& |\gq|^{\frac{b}{2}}\Big( |\gp|^2 -\frac{b^2}{4}|\gq|^{-2} \Big)   |\gq|^{\frac{b}{2}}\\
\geq& |\gq|^{b\over2} \cH_{2,n} |\gq|^{b\over2}>0,
\end{split}
\end{equation}
because $b\leq n-2$.  Since the first inequality is actually an
equality in the case $b=n-2$, it shows that our assumption $a+b\leq n$
is critical, since Herbst's inequalities are sharp.
\end{proof}

\section{Ground State Representation\label{ground1}}

The result of the previous section can be viewed as a warmup for the
following result.
\begin{theorem}
  Assume $a,b\in(0,\infty)$ with $a+b\leq n$, $\min\{a,b\}\in(0,2)$, and $\psi\in C^\infty_0(\rz^n)$. Then
  \begin{multline}
    \label{li-formula}
    (\psi,(\cJ_{a,b,n}-L_{a,b,n}|\gq|^{b-a})\psi)
=(\psi,|\gq|^{b\over2}\cH_{a,n}|\gq|^{b\over2}\psi) \\+ \alpha_{a,n} \int_{\rz^n}\rd x \int_{\rz^n}\rd y {(|x|^{b\over2}-|y|^{b\over2})^2 \left|\psi(x)|x|^\gamma-\psi(y)|y|^\gamma\right|^2\over
      2|x|^\gamma|x-y|^{n+a}|y|^\gamma}
  \end{multline}
  where $\gamma= (n+b-a)/2$ and
  \begin{equation}
    \label{eq:liconstant}
    L_{a,b,n}= 2^a {\Gamma({n-b+a\over4})\Gamma({n+b+a)\over4})\over \Gamma({n+b-a\over4})\Gamma({n-b-a\over4})}.
  \end{equation}
\end{theorem}
\begin{description}
\item[Monotony of $L_{a,b,n}$] Note that $L_{a,b,n}$ is a
  strictly monotone decreasing function in $b$ on the interval
  $[0,n-a]$ and vanishes at $n-a$. The second claim is obvious, since
  $\lim_{x\to0_+}\Gamma(x)=0$. For the first claim we use the log
  convexity of the Gamma function (Bohr and Mollerup).
\item[Sharpness] Formula \eqref{li-formula} implies the inequality
  \begin{equation}
    \label{streng}
    \cJ_{a,b,n} > L_{a,b,n}|\gq|^{b-a} + |\gq|^{b/2}\cH_{a,n}|\gq|^{b/2}
  \end{equation}
  is strict under the assumptions of the theorem, since the remainder
  term in \eqref{li-formula} vanishes, if and only if
  $\psi(x)=c|x|^{-\gamma}$ which is only in $L^2$ when $c=0$. However,
  the remainder can be made arbitrarily small by a smooth cut-off
  tending to infinity.

  If $a=2$, equality holds in \eqref{streng} because of the
  calculation \eqref{giveanumber}.
\end{description}
\begin{proof}
  Pick $\gamma:=\frac{n+b-a}{2}$. By Fourier transform of
  $|\cdot|^{-\alpha}$ (see \eqref{eq:fourier}), we know that
  \begin{eqnarray}
    \label{FourierJ}&&(|\psi|^2|x|^\gamma, \cJ_{a,b,n} |x|^{-\gamma})\\
    \nonumber &=& \dfrac{1}{2}\int_{\rz^n} \rd\xi |\xi|^a \Big(\overline{(|\psi(x)|^2|x|^\gamma)^{\wedge}(\xi)} (|x|^{b-\gamma})^{\wedge}(\xi) \\
    \nonumber && \qquad +\overline{(|\psi(x)|^2|x|^{\gamma+b})^{\wedge}(\xi)} (|x|^{-\gamma})^{\wedge}(\xi)\Big)\\
    \nonumber &=& \dfrac{1}{2}\Big(\dfrac{B_{n-(\gamma-b)}}{B_{\gamma-b}}\int_{\rz^n} \rd\xi |\xi|^{a-n+(\gamma-b)} (|\psi(x)|^2|x|^\gamma)^{\wedge}(\xi) \\ \nonumber && \qquad +\dfrac{B_{n-\gamma}}{B_\gamma} \int_{\rz^n}\rd\xi |\xi|^{a-n+\gamma}(|\psi(x)|^2|x|^{\gamma+b})^{\wedge}(\xi)\Big)\\
    \nonumber &=& \dfrac{1}{2}\Big(\dfrac{B_{a+\gamma-b}B_{n-(\gamma-b)}}{B_{n-a-(\gamma-b)}B_{\gamma-b}} +\dfrac{B_{a+\gamma}B_{n-\gamma}}{B_{n-a-\gamma}B_\gamma} \Big)\int_{\rz^n}\rd x |\psi(x)|^2|x|^{b-a}\\
    \nonumber &=& 2^a {\Gamma({n-b+a\over4})\Gamma({n+b+a)\over4})\over \Gamma({n+b-a\over4})\Gamma({n-b-a\over4})}\int_{\rz^n}\rd x |\psi(x)|^2|x|^{b-a}\\
    \nonumber &=& L_{a,b,n}\int_{\rz^n}\rd x |\psi(x)|^2|x|^{b-a}.
\end{eqnarray}
(Note that we refrained from doing obvious mollifications.) We have a
similar computation for the operator
$|\gq|^{\frac{b}{2}}|\gp|^a|\gq|^{\frac{b}{2}}$,
\begin{eqnarray}
  \label{FourierH}&&(|\psi|^2|x|^\gamma, |\gq|^{\frac{b}{2}}|\gp|^a|\gq|^{\frac{b}{2}} |x|^{-\gamma})\\
  \nonumber &=& \int_{\rz^n} \rd\xi |\xi|^a \overline{(|\psi(x)|^2|x|^{\frac{b}{2}+\gamma})^{\wedge}(\xi)} (|x|^{\frac{b}{2}-\gamma})^{\wedge}(\xi) \\
  \nonumber &=& \dfrac{B_{n-(\gamma-\frac{b}{2})}}{B_{\gamma-\frac{b}{2}}}\int_{\rz^n} \rd\xi |\xi|^{a-n+(\gamma-\frac{b}{2})} (|\psi(x)|^2|x|^{\frac{b}{2}+\gamma})^{\wedge}(\xi) \\ 
  \nonumber &=& \dfrac{B_{a+\gamma-\frac{b}{2}}B_{n-(\gamma-\frac{b}{2})}}{B_{n-a-(\gamma-\frac{b}{2})}B_{\gamma-\frac{b}{2}}} \int_{\rz^n}\rd x |\psi(x)|^2|x|^{b-a}\\
  \nonumber &=& 2^a \left({\Gamma({n+a\over4})\over \Gamma({n-a\over4})}\right)^2\int_{\rz^n}\rd x |\psi(x)|^2|x|^{b-a}.
\end{eqnarray}
On the other hand, by using \eqref{eq:betrag} and polarization we can
compute the above quantities again and obtain
\begin{eqnarray}
  \label{BetragJ}&&(|\psi|^2|x|^\gamma, \cJ_{a,b,n} |x|^{-\gamma})\\
  \nonumber &=& \dfrac{1}{2}\alpha_{a,n}\int_{\rz^n} \int_{\rz^n} \dfrac{\rd x\rd y}{|x-y|^{n+a}} \Big\{ (|\psi(x)|^2|x|^\gamma -|\psi(y)|^2|y|^\gamma) (|x|^{b-\gamma}-|y|^{b-\gamma})\\
  \nonumber && \qquad +(|\psi(x)|^2|x|^{b+\gamma} -|\psi(y)|^2|y|^{b+\gamma}) (|x|^{-\gamma}-|y|^{-\gamma})\Big\}\\
  \nonumber &=& \dfrac{1}{2}\alpha_{a,n}\int_{\rz^n} \int_{\rz^n} \dfrac{\rd x\rd y}{|x-y|^{n+a}} \Big\{ 2|\psi(x)|^2|x|^b +2|\psi(y)|^2|y|^b\\
  \nonumber && \qquad -|\psi(x)|^2|x|^\gamma|y|^{b-\gamma} -|\psi(y)|^2|x|^{b-\gamma}|y|^{\gamma}\\
  \nonumber && \qquad -|\psi(x)|^2 |x|^{b+\gamma}|y|^{-\gamma} -|\psi(y)|^2|x|^{-\gamma}|y|^{b+\gamma}\Big\}.
\end{eqnarray}
By \eqref{eq:4} and subtraction and addition of $2\Re \overline{\psi(x)}
\psi(y) |y|^b +2\Re \psi(x)\overline{\psi(y)} |x|^b$ in the above
braces we get
\begin{eqnarray}
  \label{BetragJmore}&&(|\psi|^2|x|^\gamma, \cJ_{a,b,n} |x|^{-\gamma})\\
  \nonumber &=& \alpha_{a,n}\Re\int_{\rz^n} \int_{\rz^n} \dfrac{\rd x\rd y}{|x-y|^{n+a}}(\overline{\psi(x)}-\overline{\psi(y)})(|x|^b\psi(x)-|y|^b\psi(y))\\
  \nonumber && \qquad +\dfrac{1}{2}\alpha_{a,n}\int_{\rz^n} \int_{\rz^n} \dfrac{\rd x\rd y}{|x-y|^{n+a}} \Big\{ 2\Re \overline{\psi(x)} \psi(y) |y|^b +2\Re \psi(x)\overline{\psi(y)} |x|^b\\
  \nonumber && \qquad \qquad -|\psi(x)|^2|x|^\gamma|y|^{b-\gamma} -|\psi(y)|^2|x|^{b-\gamma}|y|^{\gamma}\\
  \nonumber && \qquad \qquad -|\psi(x)|^2 |x|^{b+\gamma}|y|^{-\gamma} -|\psi(y)|^2|x|^{-\gamma}|y|^{b+\gamma}\Big\}\\
  \nonumber &=& (\psi, \cJ_{a,b,n}\psi)- \alpha_{a,n}\int_{\rz^n} \int_{\rz^n} \dfrac{\frac{1}{2}(|x|^b+|y|^b)\rd x\rd y}{|x|^\gamma|x-y|^{n+a}|y|^\gamma}\big|\psi(x)|x|^\gamma -\psi(y) |y|^\gamma\big|^2.
\end{eqnarray}
The sesquilinear form of
$|\gq|^{\frac{b}{2}}|\gp|^a|\gq|^{\frac{b}{2}}$ is
\begin{eqnarray}
  \label{BetragH}&&(|\psi|^2|x|^\gamma, |\gq|^{\frac{b}{2}}|\gp|^a|\gq|^{\frac{b}{2}} |x|^{-\gamma})\\
  \nonumber &=& \alpha_{a,n}\int_{\rz^n} \int_{\rz^n} \dfrac{\rd x\rd y}{|x-y|^{n+a}}(|\psi(x)|^2|x|^{\frac{b}{2}+\gamma}-|\psi(y)|^2|y|^{\frac{b}{2}+\gamma})(|x|^{\frac{b}{2}-\gamma}-|y|^{\frac{b}{2}-\gamma})\\
  \nonumber &=& \alpha_{a,n}\int_{\rz^n} \int_{\rz^n} \dfrac{\rd x\rd y}{|x-y|^{n+a}}\Big\{|\psi(x)|^2|x|^b+|\psi(y)|^2|y|^b\\
  \nonumber && \qquad\qquad -2\Re\overline{\psi(x)}\psi(y)|x|^{\frac{b}{2}}|y|^{\frac{b}{2}} +2\Re\overline{\psi(x)}\psi(y)|x|^{\frac{b}{2}}|y|^{\frac{b}{2}}\\
  \nonumber && \qquad\qquad -|\psi(x)|^2|x|^{\frac{b}{2}+\gamma}|y|^{\frac{b}{2}-\gamma} -|\psi(y)|^2 |x|^{\frac{b}{2}-\gamma}|y|^{\frac{b}{2}+\gamma}\Big\}\\
  \nonumber &=& (\psi, |\gq|^{\frac{b}{2}}|\gp|^a|\gq|^{\frac{b}{2}} \psi) -\alpha_{a,n} \int_{\rz^n} \int_{\rz^n} \rd x\rd y \dfrac{\big|\psi(x)|x|^\gamma -\psi(y)|y|^\gamma\big|^2}{|x|^{\gamma-\frac{b}{2}}|x-y|^{n+a}|y|^{\gamma-\frac{b}{2}}}.
\end{eqnarray}
A combination of the computations \eqref{FourierJ} to \eqref{BetragH}
and the ground state representation
\eqref{eq:groundstatetransformedhardy}gives us the desired result.
\end{proof}

\appendix
\section{Auxiliary Facts\label{a1}}
For the reader's convenience we collect some helpful known facts:
\begin{description}
\item[Fourier transforms of powers]
 For $\alpha\in(0,n)$ 
  \begin{equation}
    \label{eq:fourier}
    B_\alpha\cF(|\cdot|^{-\alpha}) = B_{n-\alpha}|\cdot|^{-n+\alpha}
  \end{equation}
  with $B_\alpha:=2^{\alpha\over2}\Gamma(\alpha/2)$ (see, e.g., Lieb
  and Loss \cite[Theorem 5.9]{LiebLoss1996})
  \item[Generalized Hardy Inequalities (Herbst \cite{Herbst1977})]
  Assume $a\in(0,n)$. Then, on $H^{a/2}(\rz^n)$
  \begin{equation}
    \label{eq:hardy}
    \cH_{a,n}:=|\gp|^a -2^a\left[{\Gamma\left({n+a\over4}\right)\over\Gamma\left({n-a\over4}\right)}\right]^2 |\gq|^{-a}>0.
  \end{equation}
  The inequality is sharp in the sense that there is no smaller
  constant in front of $|\gq|^{-a}$ which allows this inequality on
  $C^\infty_0(\rz^n)$.

Hardy's classical inequality is obtained for $a=2$, Kato's inequality
is the case $a=1$.
\item[Fractional Laplacian]
For $\psi\in H^{a/2}(\rz^n)$
 \begin{equation}
    \label{eq:betrag}
    (\psi,|\gp|^a\psi) = \alpha_{a,n} \int_{\rz^n}\rd x \int_{\rz^n}\rd y
    {|\psi(x)-\psi(y)|^2\over |x-y|^{n+a}}
  \end{equation}
  with
  $$\alpha_{a,n}={2^{a-1}\over \pi^{n/2}} {\Gamma(\frac{n+a}2)\over |\Gamma(-\frac a2)|} $$
(Frank et al \cite[Formula (3.2)]{Franketal2008H}).
\item[Ground State transformed Hardy Operator (Frank et
  al \cite{Franketal2008H})] 
  For all $\psi\in C^\infty_0(\rz^n\setminus\{0\})$
  \begin{equation}
    \label{eq:groundstatetransformedhardy}
    (\psi,\cH_{a,n}\psi)= \alpha_{a,n} \int_{\rz^n}\rd x \int_{\rz^n}\rd y {\left|\psi(x)|x|^{n-a\over2}-\psi(y)|y|^{n-a\over2}\right|^2\over |x|^{n-a\over2}|x-y|^{n+a}|y|^{n-a\over2}}.
  \end{equation}
\end{description}
\textit{Acknowledgment:} We thank Rupert Frank for his interest in
this generalization of Lieb's original inequality and some helpful
remarks on the first draft of the manuscript. Further thanks go to Ira
Herbst, Kenji Yajima, and Daisuke Aiba for careful reading of the
preprint.  This work has been partially supported by the Sino-German
program ``Analysis of Partial Differential Equations and Their
Applications'' (National Science Foundation of China and Deutsche
Forschungsgemeinschaft), by the SFB 12 of the DFG, and NSFC grant
11271218 . H. S. thanks also the Mittag-Leffler Institute where part
of this work has been done.

%\bibliographystyle{plain}
%\bibliography{coulomb}

\end{document}